\documentclass[prd,aps,showpacs]{revtex4}
\usepackage{graphicx}
\begin{document}

\title{Transport of Cosmic Rays in Chaotic Magnetic Fields}
\author{Fabien Casse\footnote{Now at FOM-Institute for Plasma physics,
Postbus 1207 NL-3430 BE Nieuwegein, Netherlands. E-mail fcasse@rijnh.nl}} \affiliation{Laboratoire d'AstrOphysique de Grenoble,BP
    53 F-38041 Grenoble Cedex 9, France}
\author{Martin Lemoine} \affiliation{Institut d'Astrophysique de Paris,
CNRS, 98 bis Boulevard Arago, F-75014 Paris, France}
\author{Guy Pelletier\footnote{Institut Universitaire de France}} \affiliation{Laboratoire d'AstrOphysique de Grenoble, BP
    53 F-38041 Grenoble Cedex 9, France}
\date{\today} 
\begin{abstract}
    The transport of charged particles in disorganised magnetic fields
    is an important issue which concerns the propagation of cosmic rays
    of all energies in a variety of astrophysical environments, such as
    the interplanetary, interstellar and even extra-galactic media, as
    well as the efficiency of Fermi acceleration processes.  We have
    performed detailed numerical experiments using Monte-Carlo
    simulations of particle propagation in stochastic magnetic fields
    in order to measure the parallel and transverse spatial diffusion
    coefficients and the pitch angle scattering time as a function of
    rigidity and strength of the turbulent magnetic component. We
    confirm the extrapolation to high turbulence levels of the scaling
    predicted by the quasi-linear approximation for the scattering
    frequency and parallel diffusion coefficient at low rigidity. We
    show that the widely used Bohm diffusion coefficient does not
    provide a satisfactory approximation to diffusion even in the
    extreme case where the mean field vanishes. We find that diffusion
    also takes place for particles with Larmor radii larger than the
    coherence length of the turbulence. We argue that transverse
    diffusion is much more effective than predicted by the quasi-linear
    approximation, and appears compatible with chaotic magnetic
    diffusion of the field lines. We provide numerical estimates of the
    Kolmogorov length and magnetic line diffusion coefficient as a
    function of the level of turbulence. Finally we comment on
    applications of our results to astrophysical turbulence and the
    acceleration of high energy cosmic rays in supernovae remnants, in
    super-bubbles, and in jets and hot spots of powerful
    radio-galaxies.
\end{abstract}
\maketitle

\section{Introduction}\label{sec.intro}

The knowledge of the transport properties of charged particles in
turbulent magnetized plasmas is a long--standing problem, which bears
directly on many astrophysical issues, such as the penetration of
low-energy cosmic rays in the heliosphere~\cite{fill}, the propagation
and escape of galactic cosmic rays in and out of the interstellar
magnetic field~\cite{ptus74,berez,chan}, or even the efficiency of
Fermi acceleration mechanisms, in particular at shocks~\cite{berez}.
The diffusion coefficient transverse to the mean component of the
magnetic field plays a particularly important role in these issues,
but to date, there is no satisfactory description of perpendicular
transport. Some studies have built upon or tried to extend the results
of the ``quasi-linear theory''~\cite{jok}, whose validity is limited
to very low level turbulence, {\it i.e.} a turbulent component much
weaker than the uniform magnetic field, and which calculates the
transport coefficients by statistical averages of the displacements
perturbed to first order in the inhomogenous field. Other studies have
appealed to phenomenological approximations such as the Bohm estimate
for the diffusion coefficient $D\sim r_{\rm L}v$, which corresponds to
the assumption that the mean free path for scattering $D/v$ of a
particle of velocity $v$, is given by the Larmor radius $r_{\rm
L}$. This approximation originates from laboratory experiments which
led D. Bohm to the empirical formula $D_{\rm B} \simeq 0.06 eT/B$ for
a plasma with temperature $T$. A theoretical derivation of this
formula was proposed later by Taylor and McNamara~\cite{taylor}, and
then extended to relativistic particles~\cite{rosso}, but no theory of
Bohm diffusion (relativistic or not) in magnetic irregularities has
been derived {\it stricto-sensu} so far. Therefore it appears that
important physical and astrophysical issues are yet to be answered:

\begin{itemize}
      \item How do the transport properties change when the level of
      magnetic turbulence is increased? What are the transport
      properties when the mean field vanishes?  Notably, what is the
      relevance of the Bohm scaling?

      \item Even for low level turbulence, transverse space diffusion
      is not well known. It nevertheless plays a crucial role in the
      confinement of cosmic rays in galaxies or other extragalactic
      objects (notably radio-galaxies jets). Its magnitude is also of
      direct relevance to the performance of Fermi acceleration at
      perpendicular shocks.

      \item Do subdiffusive and more generally anomalous diffusion
      regimes exist? If yes, they are also of importance for Fermi
      acceleration.
\end{itemize}

    In order to shed light on these issues, we have performed extensive
numerical experiments to determine the pitch angle scattering rate,
and the parallel and perpendicular spatial diffusion coefficients for
a wide range of rigidities and turbulence levels. Our experiments are
conducted by Monte-Carlo simulations in which we follow the
propagation of a relativistic test particle in a stochastic magnetic
field constructed from three-dimensional Kolmogorov turbulence, and
calculate the diffusion coefficients from the statistical correlations
along the trajectory.  Our study is similar to the recent work of
Giacalone and Jokipii~\cite{GJ99} in which the spatial diffusion
coefficients in two-- and three--dimensional magnetostatic turbulence
were measured using Monte-Carlo simulations for various turbulence
levels and rigidities. Our study is however more extensive than that
of Ref.~\cite{GJ99}. In particular we measure the diffusion
coefficients in a broader range of rigidities, by studying the
diffusion of particles with Larmor radii larger than the coherence
length, and in a broader range of turbulence levels, by going up to
pure turbulence in which there is no uniform component of the magnetic
field. In contrast, Ref.~\cite{GJ99} studies the case of lower-energy
particles and smaller turbulence levels, with a turbulent magnetic
field never exceeding the uniform component in strength. We also study
in detail the pitch angle scattering rate, which is of central
interest in applications to shock acceleration processes, and study in
more detail the issue of transverse diffusion and its relation to the
chaotic wandering of field lines. Finally we will repeatedly compare
our results to Ref.~\cite{GJ99} where there is overlap, which is
important since these numerical experiments are delicate.

   Among our results, we confirm the extrapolation to high turbulence
levels of the scalings predicted by the quasi-linear theory for the
scattering rate and the parallel diffusion coefficient at low enough
rigidity. The perpendicular diffusion coefficient is shown to follow a
law which is quite different from the predictions of the quasi-linear
theory at low rigidities. We argue that its behavior is compatible
with chaotic wandering and diffusion of the magnetic field lines to
which particles are ``attached''. In particular, we demonstrate the
chaotic behavior of the magnetic field lines and calculate the
associated Kolmogorov length and diffusion coefficient in terms of the
turbulence level. We also show that the Bohm diffusion coefficient
only holds in a limited range of rigidities $0.1\lesssim \rho
\lesssim1$ for pure turbulence, and does not exist when the mean field
is non-vanishing. In this latter case the Bohm value for the
coefficient is only obtained at maximum pitch angle scattering, {\it
i.e.} for particules with Larmor radius of order of the coherence
scale.  Finally, we also found that diffusion operates even for
particles whose Larmor radii is larger than the coherence length, as
far as we have searched in rigidity (1.5 decade). On these scales, the
scattering rate decays as expected, albeit moderately, as the power
$\simeq -7/3$ of the rigidity.

    Our study is conducted with the following main simplifying
    assumptions:
\begin{itemize}
      \item The magnetic field is composed of a mean homogeneous field
      ${\bf B}_0$ and an inhomogeneous component ${\bf B}$: $\bar{\bf B}
      = {\bf B}_0 + {\bf B}({\bf x})$.

      \item The magnetic disturbances are considered to be
      static. This assumption is well justified as the waves propagate
      with velocities of the order of the Alfv\'en velocity $v_{\rm
      A}$, smaller than the velocity of particles $\sim c$ (we
      consider relativistic particles), and the electric force is thus
      smaller than the magnetic force by a factor $v_{A}/c$.  The
      first correction to the theory is the celebrated second order
      Fermi process which can be described by diffusion in momentum
      space, with diffusion coefficient $\Gamma(p) \sim \nu_{\rm s}p^2
      v_{\rm A}^2/c^2$, with $p$ particle momentum, and $\nu_{\rm s}$,
      the angular scattering frequency, is an outcome of our study.

      \item The magnetic perturbations are distributed according to
      isotropic turbulence, whose power spectrum is written in terms
      of Fourier momentum $k$ as: $\langle B(k)^2 \rangle \propto
      k^{-\beta-2}$ for $k_{\rm min}\leq k \leq k_{\rm max}$, zero
      otherwise, and $\langle {\bf B}(k) \rangle =0$, {\it i.e.}
      random phases. The exponent $\beta$ characterizes the properties
      of the turbulence, and we will concentrate on the case
      $\beta=5/3$ in our numerical applications, which describes
      Kolmogorov turbulence.  The smallest turbulence wavenumber is
      related to the maximum scale of the turbulence $L_{\rm max}$
      {\it via}: $k_{\rm min} = 2\pi/L_{\rm max}$. This largest scale
      also corresponds to the correlation length of the magnetic field
      to within a factor of order unity [see
      Eq.(\ref{eq_correlation})].

\end{itemize}

Our notations are as follows. The quantities we will be interested in
are the scattering rate $\nu_{s}$ or scattering time $\tau_{\rm s}\equiv
1/\nu_{\rm s}$, defined as the correlation time of the pitch angle, the
spatial diffusion coefficient along the mean field $D_{\parallel}$ and
the transverse spatial diffusion coefficient $D_{\perp}$. These
coefficients are evaluated in terms of turbulence level $\eta\equiv
\langle  {\bf B}^2\rangle /\langle\bar B^2\rangle=\langle{\bf B}^2\rangle
/[B^2_o+\langle {\bf B}^2\rangle]$, and rigidity $\rho\equiv 2\pi
r_{\rm L}/L_{\rm max}=r_{\rm L}k_{\rm min}$. For convenience, the
Larmor radius $r_{\rm L}$ is defined with respect to the total
magnetic field: $r_{\rm L}\equiv \epsilon/Ze\bar{B}$ for a particle
with energy $\epsilon$ and charge $Ze$. The Larmor pulsation of a
particle of energy $\epsilon$ is defined, for convenience, as $\bar
\omega_{\rm L} \equiv Ze\bar B c/\epsilon$, and the Larmor time
$t_{\rm L} \equiv (\bar \omega_{\rm L})^{-1}$. We define the
scattering function as $g(\rho, \eta) \equiv \nu_{\rm s}/\bar
\omega_{\rm L}=t_{\rm L}/\tau_{\rm s}$. When useful, we will denote by
$\omega_{\rm L}$ the Larmor pulsation in the mean field.

The paper is organised as follows. In Section 2, we recall the
relation between spatial diffusion and the scattering off magnetic
disturbances and present the numerical method. In Section 3, we
present our numerical results and discuss the issue of transverse
diffusion and the measurement of magnetic chaos characteristics.  A
discussion with direct astrophysical consequences of our results is
given in Section 5, and conclusions are offered in Section 6. Finally,
in Appendix~A, we propose a theoretical interpretation of the regimes
of diffusion observed, and in particular of the existence of diffusion
for Larmor radii larger than the maximum scale of turbulence.

\section{Momentum scattering and spatial diffusion}\label{sec.scat}

High energy particles interact with cosmic matter mostly through
scattering on the magnetic field which is more or less frozen in the
medium. The interaction is elastic in the frame of a magnetic
disturbance and it can be considered as elastic in the plasma rest
frame to lowest order in $v_{\rm A}/c$, if the disturbance propagates
at small enough velocity $v_{\rm A}\ll c$. With respect to a given
direction, chosen as that of the uniform component of the magnetic
field if this latter is non-vanishing, the pitch angle of the particle
changes almost randomly if the magnetic field is sufficiently
disorganized (this will be made more precise further on). Therefore
the position of the particle changes according to a random walk on a
time scale which is longer than the coherence time of the pitch angle
cosine, and pitch angle scattering is thus responsible for the
diffusion of particles. However it is generally believed that
transverse diffusion may also occur through wandering of the magnetic
field lines. In this picture, the transverse velocity of the particle
changes through resonant diffusion as before while the guiding center
of the approximate helical motion wanders with the magnetic field line
to which it is attached, and performs a random walk in the transverse
direction. These notions will be quantified in the forthcoming
sections. Our main objective here is indeed to quantify these various
contributions to the process of diffusion.

\subsection{Definitions -- Scattering time and diffusion
coefficients}\label{subsec.scapar}

We define the pitch angle $\alpha$ with respect to the mean field
direction when it exists, otherwise the direction can be arbitrarily
chosen.  The convenient random function is the pitch angle cosine:
$\mu(t) \equiv \cos\left(\alpha\right)$, and $\alpha$ is a function of
time. Since we assume a static spectrum of magnetic perturbations, the
auto-correlation function of $\mu(t)$ will become stationary in the
large time limit. It can then be defined as:

\begin{equation}
      C(\tau) \equiv \langle  \mu(t+\tau) \mu(t)\rangle  /
                       \langle  \mu(t)^2\rangle  \,
      \label{eq:FCMU}
\end{equation}
where the average can be performed in three different ways. In the original
quasi-linear theory, this average is taken over the phases of
the magnetic disturbances. In the theory of chaos, the average is performed
over the phase space subset of chaotic motions. In practice, and this is
what we will use in the numerical experiment, we assume ergodicity and make
temporal average. Our procedure of calculating averages is explicited
further below.

   The scattering time $\tau_{\rm s}$ can then be defined as the coherence
time of the pitch angle cosine:
\begin{equation}
      \tau_{\rm s} \equiv \int_{0}^{\infty}{\rm d}\tau\, C(\tau) \ .
      \label{eq:TAU}
\end{equation}
In particular, if the auto-correlation function falls off
exponentially $C(\tau)=\exp\left(-\tau/T\right)$ then $\tau_{\rm
s}=T$. Turning to the spatial diffusion coefficient $D_{\parallel}$,
let $x_{\parallel}$ be the coordinate of a particle along the mean
field direction. Then ${\rm d} x_{\parallel} = v \mu(t){\rm d}t$ with
a constant velocity $v$ (in our case $v=c$), since energy is
conserved. Consider now a random variation $\Delta x_{\parallel}$ of
$x_{\parallel}$ during the time interval $\Delta t$ supposed to be
larger than the scattering time $\tau_{\rm s}$. One has $\langle
\Delta x_{\parallel}\rangle \approx0$, and
\begin{equation}
      \langle  \Delta x_{\parallel}^2\rangle = v^2 \int_{t}^{t+\Delta t}
      {\rm d}t_{1} \int_{t}^{t+\Delta t} {\rm d}t_{2}\, \langle
      \mu(t_{1})\mu(t_{2})\rangle  .
      \label{eq:DXPC}
\end{equation}

Beyond the scattering time $\tau_{\rm s}$, if the stationary random process
$\mu(t)$ explores uniformly the interval $(-1,+1)$, the space
diffusion coefficient parallel to the mean field stems
straightforwardly from its definition, Eq.~(\ref{eq:DXPC}) and
Eq.~(\ref{eq:TAU}):
\begin{equation}
      D_{\parallel} \equiv \frac{\langle   \Delta x_{\parallel}^2\rangle }
       {2\Delta t} = \frac{1}{3}v^2 \tau_{\rm s} \ .  \label{eq:DPAR}
\end{equation}
Here as well the average can be made according to one of the three ways
explained above. The main goal of the computation is then to determine
the dependence of $\tau_{\rm s}$, or the scattering function $g$, in terms
of the rigidity $\rho$ and the turbulence level $\eta$. The
theoretical result is known in the regime of weak
turbulence~\cite{jok}, if the correlation time $\tau_{c}$ of the force
suffered by the particle is much smaller than the scattering time
$\tau_{\rm s}$. To make it more precise, particles undergo resonances with
the MHD modes such that $k_{\parallel}v\mu\pm\omega_{\rm L}=0$. The
correlation time is related to the width of the resonance in the mode
spectrum, such that :
\begin{equation}
\tau^{-1}_{c}=\Delta (k_{\parallel}v\mu\pm\omega_{\rm L})=v|\mu|\Delta
k_{\parallel} = \omega_{\rm L}\frac{\Delta k_{\parallel}}{k_{\parallel}}
\end{equation}
   where $\Delta k_{\parallel}$ denotes the spectrum width, in the
   parallel direction. Since $\tau^{-1}_s\sim \eta\omega_{\rm L}$,
   $\tau_{c}\ll \tau_{\rm s}$ is equivalent to $\eta \ll \Delta
   k_{\parallel}/k_{\parallel}$. In this case the memory of the initial
   pitch angle can even be kept and the scattering function $g \sim \eta
   (\rho \vert \mu \vert)^{\beta -1}$. However diffusion coefficients
   calculated on timescales larger than $\tau_{\rm s}$ must be averaged over
   $\mu$.

Due to rotation invariance around the mean field direction, there is a
single transverse diffusion coefficient (when diffusion occurs), given by:

\begin{equation}
      D_{\perp} \equiv \frac{\langle  \Delta x_{\perp}^2\rangle }{2\Delta
      t}, \label{eq:DDP'}
\end{equation}
where $\Delta x_\perp$ denotes the displacement perpendicular to the
mean field during the time interval $\Delta t$.  In weak turbulence
theory ($\eta \ll 1$), the gyro-phase $\psi$ of the particle is only
weakly perturbed by the disorganized component of the field and $\dot
\psi \simeq \omega_{\rm L}$, where the gyro-pulsation $\omega_{\rm L}$ is
determined with respect to the mean field. The transverse velocity can
be approximated by:
\begin{equation}
      {\bf v}_{\perp} \simeq v \sin \alpha(t)\left[{\bf e}_{1}\cos \psi
       - {\rm sign}(q) {\bf e}_{2}\sin \psi\right] , \label{eq:VPERP}
\end{equation}
where $q$ denotes the charge of the particle, and we implicitly
assumed the mean field to lie along the direction ${\bf e}_3$.  The
pitch angle sine $\sin \alpha (t)$ varies on the timescale $\tau_{\rm s}$,
which is much longer than the Larmor time in the weak turbulence
regime. The time correlation function of the pitch angle is obviously
the same as that of the cosines since $\langle   \cos
(\alpha_{1}+\alpha_{2}) \rangle  =0$, hence $\langle   \sin \alpha_{1}
\sin \alpha_{2}\rangle  = \langle  \cos \alpha_{1} \cos \alpha_{2}
\rangle $.  Therefore the transverse diffusion coefficient reads:

\begin{equation}
      D_{\perp} = \frac{1}{3} v^2 \int_{0}^{\infty}{\rm d}\tau\, C(\tau)
        \cos\left(\omega_{\rm L} \tau\right) \label{eq:DPQL}
\end{equation}

Assuming that the correlation function $C(\tau)$ decays exponentially
on the characteristic time $\tau_{\rm s}$, one finally obtains a result similar
to the so-called classical diffusion that reads

\begin{equation}
      D_{\perp} = \frac{1}{3} v^2 \frac{\tau_{\rm s}}{1 +
      (\omega_{\rm L}\tau_{\rm s})^2} \ .  \label{eq:DDP}
\end{equation}
This transverse diffusion based on pitch angle scattering only leads to the
ratio
\begin{equation}
       \frac{D_{\perp}}{D_\parallel} = \frac{1}{1 +
         \left(\lambda_{\parallel}/r_{\rm L}\right)^2}, \label{eq:CST}
\end{equation}
where $\lambda_{\parallel}\equiv3D_{\parallel}/v$ is the mean free
path of a particle along the mean magnetic field.  This relation can
also be obtained by treating the magnetic disturbances as hard sphere
scattering centers with weak or strong turbulence. It is also a result
of the study of Ref.~\cite{BM97}, which estimate phenomenological
diffusion coefficients by using well-motivated assumptions for the
velocity auto-correlation functions of the particle orbit. Finally,
since $(\omega_{\rm L}\tau_{\rm s})^2 \gg 1$ in the weak turbulence
regime, one expects $D_{\perp} \ll D_{\parallel}$ when $\eta\ll 1$.
However the transverse diffusion may turn out to be larger than
predicted by quasi-linear theory, even for moderate turbulence. In
particular note that Eq.~(\ref{eq:DPAR}) for the parallel diffusion
coefficient rests on the sole assumption that $C(\tau)$ vanishes on
timescales longer than $\tau_{\rm s}$, while the quasi-linear result
for the transverse diffusion coefficient, Eq.~(\ref{eq:DDP}), assumes
that the particle orbit is only weakly perturbed and the timescale of
variation of the pitch angle is much longer than the Larmor time, {\it
i.e.}, that the level of turbulence $\eta\ll 1$. We refer to this
result as a prediction of quasi-linear theory; it neglects the
diffusion of the guide center carrying field line and the associated
process of chaotic magnetic diffusion which has been analysed by
Jokipii and Parker~\cite{Jok69}, and to which we will come back in the
following Section.  Finally in all cases one should obtain $D_{\perp}
\rightarrow D_{\parallel}$ when $\eta \rightarrow 1$, since the mean
field vanishes in this limit and there is no prefered direction
anymore.

\subsection{Numerical simulations}\label{subsec.NUCO}

In order to evaluate the transport coefficients, we follow the
propagation of particles in stochastic magnetic fields by integrating
the standard equation of motion (Lorentz force), and measure the
statistical quantities of interest to us, namely $\nu_{\rm s}=
1/\tau_{\rm s}$, $D_{\parallel}$ and $D_{\perp}$, using the estimators
defined respectively in
Eqs.~(\ref{eq:TAU}),(\ref{eq:DPAR}),(\ref{eq:DDP'}). Strictly
speaking, the averages contained in these expressions should be taken
over the phases of the magnetic inhomogeneities. In practice however,
one may as well take these averages as follows. For a given $\Delta t$
[using the notations of Eqs.~(\ref{eq:DPAR}),(\ref{eq:DDP'})], a time
$t$ is picked at random over the trajectory, and the correlation
between the positions at times $t$ and $t+\Delta t$ is recorded; this
operation is repeated many times and the average is kept. This latter
is then further averaged over a population of particles with random
initial positions, and then over an ensemble of magnetic field
realizations with random phases. In practice, we propagate $20-50$
particles, measure the correlations at $5000-10000$ different times
along the trajectory of each particle, and use a few magnetic
realizations. This procedure allows to reach a sufficiently high
signal-to-noise ratio in the simulation for a moderate computer time,
as indeed setting up the magnetic field and propagating a particle is
much more costly than taking averages along the trajectory.
 
In principle one could as well take the average $\langle   \Delta
x^2\rangle  /\Delta t$ as the variance of the displacement at time
$\Delta t$ over a population of particles originally concentrated at
the origin, as in Ref.~\cite{GJ99}. However this method requires to
follow the trajectory of a large number of particles $\gtrsim10^3$ in
order to achieve a reasonable signal-to-noise ratio. The method we
employ, which measures the correlations along the trajectory of each
particle, before averaging over a population of particles, is less
costly in computer time (but requires much more memory). Nevertheless,
we also checked (and found) that the method which measures the
variance of the displacement gave results in agreement with our method
within the error attached to the small number of particles propagated.

The magnetic field can be constructed in two different ways which both
present pros and cons. The first method uses Fast-Fourier Transform
(FFT) algorithms to set up the magnetic field on a discrete grid in
configuration space, starting from the magnetic field defined through
its power spectrum in Fourier space, {\it i.e.}:

\begin{equation}
        {\bf B}({\bf x})\equiv \kappa \sum_{{\bf n}} {\bf e}({\bf n})
         A({\bf n})\exp\left[\frac{2i\pi{\bf n.x}}{L_{\rm max}}\right]
         .  \label{eq:B_FFT}
\end{equation}
In this equation, ${\bf n}$ is the tri-dimensional wavenumber vector,
with integer coordinates taking values between $1$ and $k_{\rm
max}/2k_{\rm min}$, ${\bf e}({\bf n})$ is a unit vector orthogonal to
${\bf n}$ (this ensures ${\bf \nabla B}=0$), $A({\bf n})$ is the
amplitude of the field component, and is defined such that $\langle
A({\bf n})\rangle  =0$ and $\langle   A({\bf n})A^\star({\bf n})\rangle  =
k^{-\beta-2}$, where the average concerns the phases of the magnetic
field. Finally, $\kappa$ is a numerical prefactor which ensures the
correct normalization of the inhomogenous magnetic component with
respect to the mean field, by using the following ergodic
approximation to $\langle   B^2 \rangle $:

\begin{equation}
      \langle   B^2 \rangle  = \frac{1}{V}\int {\rm d}{\bf x}\, {\bf
      B}^2({\bf x}),
\end{equation}
and as before $\eta = \langle   B^2\rangle /(B_0^2 + \langle
B^2\rangle )$. In practice, the field components are calculated at each
vertex ${\bf x}_i$ of a discrete grid in configuration space
beforehand. The boundary conditions are periodic with period
$L_{\rm max}$, and the fundamental cubic cell size is $L_{\rm
max}/N_g$, where $N_g$ represents the number of wavenumber modes along
one direction. One thus has: $k_{\rm max}/k_{\rm min} = L_{\rm
max}/L_{\rm min} = N_g/2$, where the factor 2 comes from the fact that
one must consider both negative and positive $k$-modes to respect the
hermiticity of ${\bf B}({\bf k})$. In our simulation, we typically use
$N_g=256$ and in some cases $N_g=512$ which gives us a dynamic range
of two orders of magnitude.

   During the propagation of particles, it is of course necessary to
know the magnetic field at any point ${\bf x}$ for the integration of
the equations of motion. Our numerical code calculates the value ${\bf
B}({\bf x})$ either by tri-linear interpolation between the known
values of the field components on the 8 vertices of the cell to which
${\bf x}$ belongs, or by taking the value of ${\bf B}$ at the vertex
closest to ${\bf x}$, which amounts to assuming a constant ${\bf B}$
in cells of size $L_{\rm max}/N_g$ centered on each vertex. While the
former method does not respect ${\bf\nabla B}=0$, the latter implies a
discontinuous magnetic field on each cubic cell face.  We will show in
the following that the results obtained by these methods differ only
when scales smaller than the cell size are concerned, as expected.

    A second algorithm for computing the magnetic field has been
proposed by Giacalone \& Jokipii~\cite{GJ99} and calculates the
magnetic field as a sum over plane wave modes (we will refer to this
algorithm as GJ). The expression defining ${\bf B}({\bf x})$ is very
similar to Eq.~(\ref{eq:B_FFT}) above, except that ${\bf n}$ needs not
have integral coordinates anymore, as Fast-Fourier Transform methods
are not used. Indeed, one does not calculate the field on a discrete
grid beforehand, but its values are calculated where and when needed
during the propagation directly from the sum over plane waves. Also
the sum is not tri-dimensional, but one-dimensional; the wavenumbers
directions are drawn at random, and the amplitude $A(k)\propto
k^{-\beta}$ to account for phase space volume. In practice, it is
convenient to have logarithmic spacing of the $k-$modes between
$k_{\rm min}$ and $k_{\rm max}$.

     One main advantage of the GJ method is that there is no restriction
in dynamic range due to memory usage, and consequently $k_{\rm
max}/k_{\rm min}$ can be as large as required. However, one is limited
in terms of computer usage time since it is expensive to perform the
sum over the wavenumber modes at each point of the trajectory if the
number of modes $N_{pw}$ becomes significant. In practice, $N_{pw}=
500$ is a strict upper limit for our applications~\cite{Gpc}, and even
with $N_{pw}=200$ the calculation is already much slower than a
similar calculation with the above FFT algorithm.

    The number of modes is important as it controls the efficiency of
diffusion, since pitch angle scattering proceeds mainly through
resonance of the particle momentum on the magnetic field modes. In
quasi-linear theory the resonance condition reads $\rho\mu
k_{\parallel}=\pm k_{\rm min}$, where $k_{\parallel}$ is the component
of the wavenumber along the mean magnetic field direction. The FFT and
GJ share a similar number of resonance modes in this limit $\eta \ll
1$. However for each resonant $k_{\parallel}$ the FFT algorithm has
$N_g^2\sim 10^4-10^5$ transverse components to be compared with one
for the GJ algorithm. One thus expects that at higher turbulence
levels, diffusion should be more effective in the FFT algorithm due to
the much larger total number of modes than in the GJ
algorithm. Furthermore, in order to preserve a correct spacing of
modes in the GJ algorithm, one cannot indefinitely increase the
dynamic range $k_{\rm max}/k_{\rm min}$ since $N_{pw}$ is fixed for
practical reasons, {\it i.e.} computer time.

   However the FFT algorithm suffers from other limitations (apart from
the limitation in dynamic range): the interpolation of ${\bf B}$ on
scales smaller than the cell size, and the periodicity on the scale
$L_{\rm max}$. These limitations are not present in the GJ algorithm,
and imply that the results of the FFT method obtained for Larmor radii
much smaller than the cell size, {\it i.e.} $\rho \ll 1/N_g$, or much
bigger than the periodicity scale, $\rho \gg 1$, cannot be trusted,
since these regimes are likely to be dominated by systematic effects
related to the discreteness or to the periodicity.

Overall both methods appear complementary to each other, and we use
them in turn to compare and discuss the robustness of our numerical
results with respect to the assumptions made.

\section{Results and discussion}\label{subsec.NURE}

\subsection{Pitch angle scattering and parallel diffusion}

\begin{figure}[t]
      \centering \includegraphics[width=10cm,clip=true]{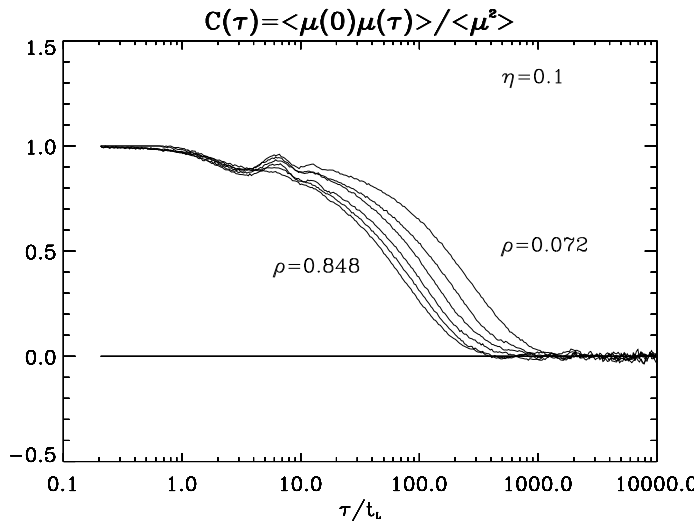}
      \caption[...]{Self-correlation function of the pitch angle cosine
      shown as a function of time $\tau$ (in units of Larmor time
      $t_{\rm L}=1/\omega_{\rm L}$) for various rigidity $\rho = 0.072, 0.12,
      0.19, 0.32, 0.52, 0.85$ and for $\eta=0.1$.}  \label{fig:corr}
\end{figure}
\begin{figure}[t]
      \centering \includegraphics[width=10cm]{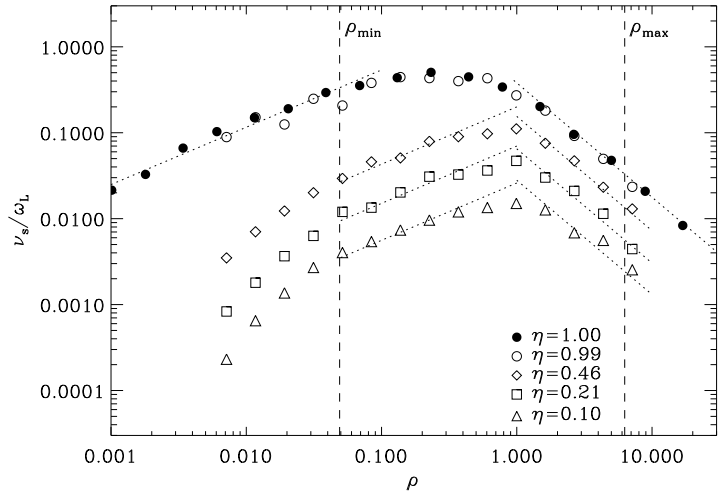}
      \caption[...]{The scattering function $g(\eta,\rho)=\nu_{\rm
      s}/\bar\omega_{\rm L}$ as a function of rigidity $\rho$. The
      symbols correspond to the measurements made through our
      Monte-Carlo experiments and correspond to various turbulence
      levels, as indicated. These results have been obtained using the
      FFT numerical method (see text), except for the filled circles
      which correspond to the GJ algorithm. The vertical dashed lines
      indicate the range of validity of our FFT algorithm ({\it i.e.}
      all symbols except crosses), delimited by $\rho_{\rm min} =
      k_{\rm min}/k_{\rm max}$, and $\rho_{\rm max} = 2\pi$, which
      correspond respectively to Larmor radii $r_{\rm L}=L_{\rm
      max}/\pi N_g$ ($1/\pi$ cell size) and $r_{\rm L}=L_{\rm
      max}$. The simulation for $\eta=1$ shown by crosses has been
      obtained with a much larger dynamic range than the others, {\it
      i.e.} $k_{\rm max}/k_{\rm min} = 10^4$. Finally, the dotted lines
      correspond to power law approximations with slopes $2/3$ and
      $-4/3$. See text for comments. } \label{fig:scatt}
\end{figure}

The first numerical investigation to perform is the self-correlation
function of the pitch angle cosine. The behavior of this function is
shown in Fig.~(\ref{fig:corr}) {\it vs} time interval $\tau$ for
various levels of turbulence $\eta$. Two bumps are observed at one and
two Larmor periods. These bumps are observed as long as the regular
magnetic field ${\bf B}_o$ exists. Since the decorrelation times for
$\eta < 1$ are larger than one Larmor period, the Larmor motions are
not completely disorganized and contribute to the correlation function
with some harmonics generated by nonlinearities. The inflexion of the
function indicates that it behaves as $e^{-a\tau^2}$ as $\tau\to 0$
and then decreases exponentially in $e^{-\nu \tau}$ as
$\tau\to\infty$. Thus the determination of $\tau_{\rm s}$ and
$\nu_{s}$ by a numerical integration of the correlation function is
accurate.

In Fig.~(\ref{fig:scatt}) we show the scattering frequency
$g(\eta,\rho)=\nu_{\rm s}/\bar\omega_{\rm L}$, which is the main
quantity of interest for evaluating the transport coefficients. This
figure shows several interesting features which deserve further
comments. First of all, one finds that both methods for calculating
the magnetic field, {\it i.e.}  FFT and GJ, agree well within the
range of validity of the former method, namely for $\rho_{\rm
min}\lesssim \rho \lesssim \rho_{\rm max}$, where $\rho_{\rm
min}=k_{\rm min}/k_{\rm max}$, and $\rho_{\rm max}=2\pi$. These two
limiting rigidities correspond to Larmor radii of order of the cell
size and of the maximum scale of turbulence respectively, and result
from the discreteness and periodicity of the magnetic field grid, as
explained in Section~\ref{subsec.NUCO}. One finds that the scattering
function behaves as a power-law with different slopes depending on the
rigidity and turbulence level. For $\rho\lesssim \rho_{\rm min}$, it
must be emphasized that the results cannot be trusted for the FFT
results, {\it i.e.} all symbols except filled circles, and the change
of slope may be artificial. For $\eta<1$ and $\rho<1$, it appears that
$g(\eta,\rho)\propto \eta\rho^{2/3}$, in accordance with the
quasi-linear prediction since $2/3=\beta-1$.

For $\rho>1$, one finds $g(\eta,\rho) \propto \eta\rho^{-4/3}$, an
unexpected result, since the resonance conditions cannot be
satistified at these high rigidities, and the quasi-linear theory thus
predicts a sudden drop of the scattering frequency at $\rho>1$. In
Appendix~A, we provide a first theoretical explanation of this result
by expansion of the particle trajectory in the random displacement and
statistical averaging of a non-perturbative resummation of an infinity
of graphs of correlations along the trajectory.

For $\eta=0.99$ and $\rho\lesssim 1$ one notes a flattening of the
scattering function with recovery of the exponent $2/3$ power law at
smaller rigidities. This flattening is definitely present for $\eta=1$
(no mean component of the magnetic field), and correspond to the
phenomenological Bohm diffusion regime, as will be seen further below;
however, it only extends over slightly less than a decade in rigidity
for $0.1\lesssim\rho\lesssim1$, even though for that simulation the
dynamic range was very large $k_{\rm max}/k_{\rm min}=10^4$. At
maximum pitch angle scattering, i.e. when $\rho\simeq1$ and
$\eta\simeq1$, the scattering function $g\simeq0.5$, {\it i.e.} the
pitch angle scattering time $\tau_{\rm s}$ is of order 2 Larmor times
$t_{\rm L}$.
It should be noted that we define the rigidity with respect to the
maximum scale of turbulence, which strictly speaking does not coincide
with the coherence scale $l_{\rm coh}$ of the turbulent magnetic
field. In effect, the spatial correlation function of the turbulent
component is defined as
\begin{equation}\label{eq_correlation}
\langle{\bf B}({\bf x}+{\bf r}){\bf B}({\bf x})\rangle =
\langle{\bf B}^2\rangle \frac{\int {\rm d}k\,
\frac{\sin(kr)}{kr}S(k)}{\int {\rm d}k\, S(k)},
\end{equation}
with $S(k)\equiv k^2\langle {\bf B}^2(k)\rangle$ the power
spectrum. This integral cannot be integrated analytically for a
power-law spectrum $S(k)\propto k^{-\beta}$ but one can check
numerically that the maximum of the correlation function occurs at
scale $l_{\rm coh}\simeq 0.77L_{\rm max}/2\pi$.

\begin{figure}[t]
      \centering \includegraphics[width=10cm]{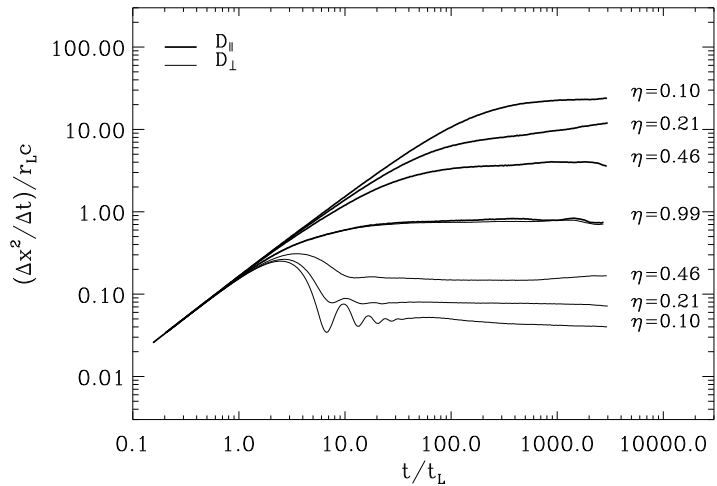}
      \caption[...]{Behavior of the averages $\langle  \Delta x^2\rangle
      /\Delta t$ in units of $r_{\rm L}c$, as a function of the time
      interval $\Delta t$ in units of $t_{\rm L}$, for various
      turbulence levels ($\rho=0.848$), and for both the transverse
      displacement (lower
      thin line curves) and parallel displacement (upper thick
      curves). One sees the transition from the weakly perturbed
      propagation regime $\langle   \Delta x^2\rangle  \propto \Delta t^2$
      to the diffusion regime $\langle   \Delta x^2\rangle  \propto \Delta
      t$, which appears here as a plateau. The transition duration
      depends on the turbulence level, and is of order of $\tau_{\rm s}$ the
      scattering time. The diffusion coefficients are given by the
      levels of the plateaux. Obviously, $D_\parallel\gg D_\perp$ for
      $\eta < 1$ and the two meet in the limit $\eta\to1$, as expected.}
      \label{fig:diff}
\end{figure}

Turning to the spatial diffusion coefficients, it is interesting to
plot the statistical estimators for $D_\parallel$ and $D_\perp$ given
by Eqs.~(\ref{eq:DPAR}),(\ref{eq:DDP'}) as a function of time for
different turbulence levels, and the result is shown in
Fig.~(\ref{fig:diff}).

This figure illustrates the transition from the regime in which the
particle orbit is weakly perturbed and memory of the initial
conditions is kept to the regime in which this memory is lost and the
particle diffuse, $\langle   \Delta x^2\rangle /\Delta t
\approx$constant. The level of this plateau gives the magnitude of the
diffusion coefficient; Fig.~(\ref{fig:diff}) also gives an idea of the
uncertainty in our measurement of diffusion coefficients.
Finally, this figure also confirms the expected results $D_\parallel
\gg D_\perp$ when $\eta\ll1$ and $D_\parallel/D_\perp \to 1$ as
$\eta\to 1$. It should be pointed out that the initial value of the
pitch angle cosine was $\mu=1/\sqrt{2}$ in all simulations; we have
checked that our results are insensitive to this value as long as the
turbulence level $\eta\gtrsim 0.1$, as expected.

\begin{figure}[htb]
      \centering \includegraphics[width=10cm]{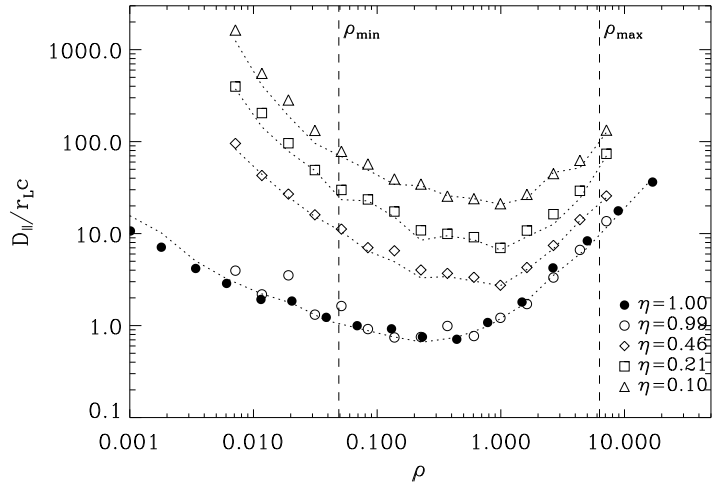}
      \caption[...]{The parallel diffusion coefficient $D_\parallel$ in
      units of $r_{\rm L}c$ as a function of rigidity for various
      turbulence levels. The symbols and vertical dashed lines are as
      in Fig.~(\ref{fig:scatt}). The dotted lines are obtained from the
      pitch angle scattering rate, using Eq.~(\ref{eq:DPAR}).  }
      \label{fig:Dpar}
\end{figure}

In Fig.~(\ref{fig:Dpar}), we show the behavior of the parallel
diffusion coefficient as a function of rigidity for various turbulence
levels. The dotted lines correspond to the approximation of
$D_\parallel$ obtained from the calculation of $\tau_{\rm s}$ using
Eq.~(\ref{eq:DPAR}), and the agreement appears excellent. This study
does not confirm the existence of a Bohm scaling. More precisely, the
Bohm diffusion coefficient $D_{\rm B}\propto r_{\rm L}v$ only applies
at $\eta=1$ in the range $0.1\lesssim\rho\lesssim1$, in agreement with
the similar conclusion for the scattering function. In all other cases
the quasi-linear prediction is verified, {\it i.e.}
$D_\parallel\propto \rho^{1/3}$ for $\rho <1$.  We also found that a
diffusion regime exists for rigidities greater than the upper bound of
the resonance region, {\it i.e.} $\rho > 1$, for as far as we have
searched, or about 1.5 decade. In this regime $\rho>1$,
$D_\parallel\propto \rho^{7/3}$, for all values of $\eta$.


\subsection{The issue of transverse diffusion}\label{sec.DPERP}

\begin{figure}[t]
      \centering \includegraphics[width=10cm]{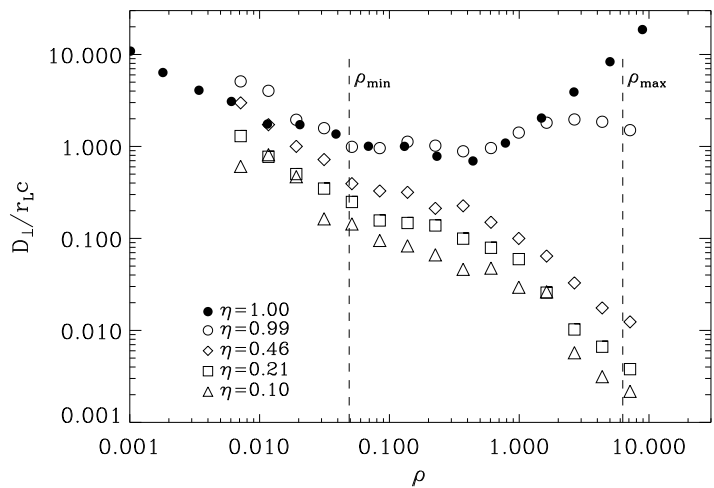}
      \caption[...]{The transverse diffusion coefficient as a function of
      rigidity for various turbulence levels, with the same notations
      for the symbols as in Fig.~(\ref{fig:Dpar}). } \label{fig:Dperp}
\end{figure}

\begin{figure}[htb]
      \centering
      \includegraphics[width=10cm,clip=true]{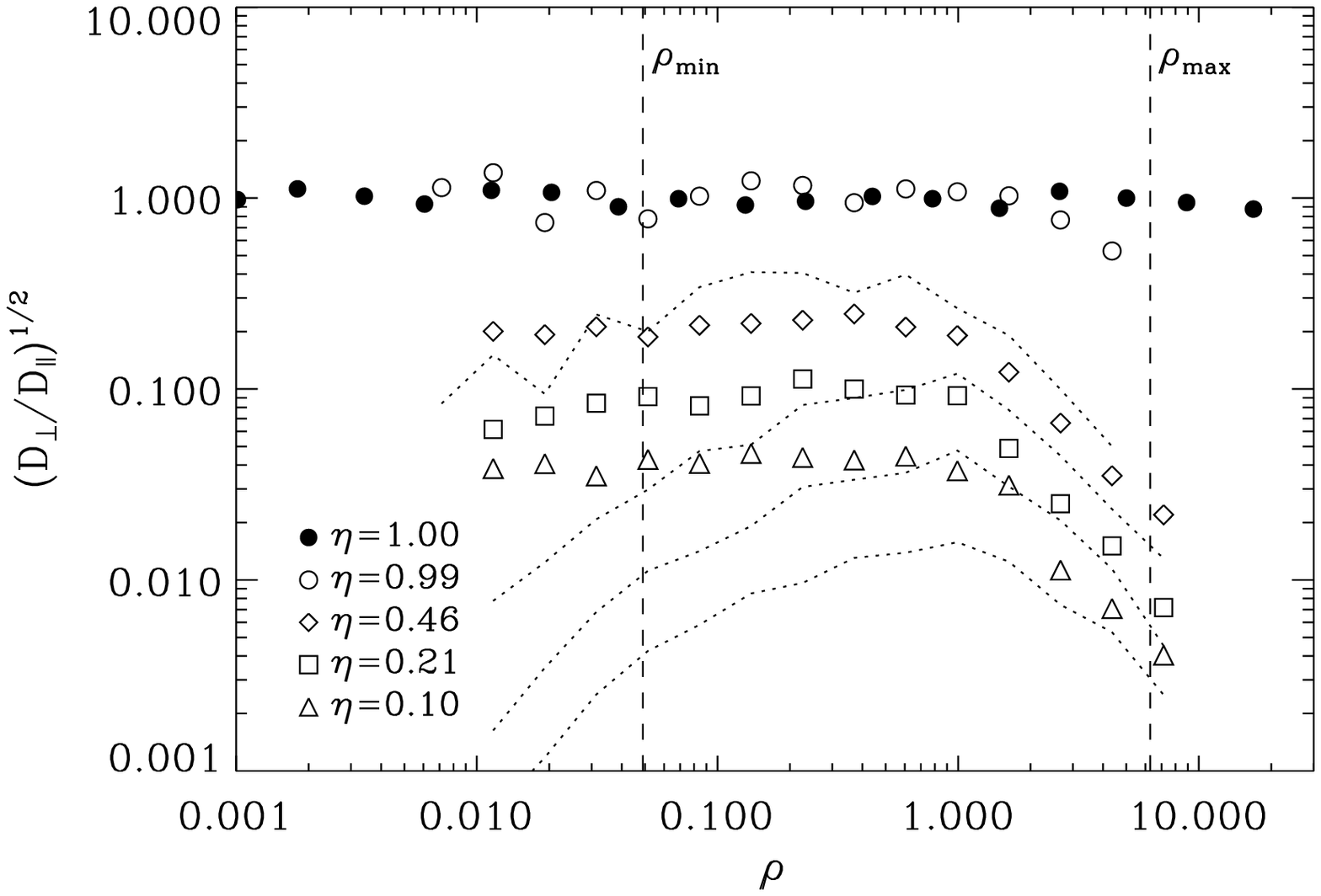}
      \caption[...]{The square root of $D_{\perp}/D_{\parallel}$ as a
      function of rigidity for various values of $\eta$. The notations
      of symbols are as indicated and as in previous figures. The
      dotted curves overlaid on this figure correspond to the classical
      scattering result given by Eq.~(\ref{eq:CST}), and correspond
      from bottom to top to the represented values of $\eta$ in
      increasing order except for $\eta=1$. These models account
      marginally for the numerical results for high rigidity particles
      $\rho\gtrsim1$ and small turbulence levels $\eta \lesssim 0.5$,
      but diverge significantly from the experiment in other regimes.}
      \label{fig:Dper2Dpar}
\end{figure}

In Fig.~(\ref{fig:Dperp}), we plot the behavior of the transverse
diffusion coefficient as a function of rigidity for various turbulence
levels.  It is useful to plot also the quantity
$\left(D_{\perp}/D_{\parallel}\right)^{1/2}$ as shown in
Fig.~(\ref{fig:Dper2Dpar}). Indeed, the noise of the simulation is
then reduced and this figure allows to compare directly the power law
behaviors of $D_\perp$ and $D_\parallel$.

This figure indeed reveals a clear trend. For all $\eta$, the ratio
$D_\perp/D_\parallel$ is independent of rigidity for $\rho<1$ , and
scales as $\rho^{-2}$ for $\rho>1$. A similar regime has been found by
Giacalone \& Jokipii~\cite{GJ99} for $\rho < 1$, albeit with slightly
lower values than ours. This constancy is interpreted in the following
as the signature of diffusion due to the chaotic wandering of the
guide center carrying field lines. The importance of the guiding
center diffusion was pointed out by Jokipii~\cite{jok} as early as
1966 in order to correct the quasi-linear result; however this
derivation does not apply to high turbulence levels. Finally, the
ratio $D_\perp/D_\parallel$ converges as expected to $1$ for all
$\rho$ when $\eta\to 1$. However it is interesting to note that even
at $\eta=0.99$, there remains the power law dependence for $\rho>1$,
$D_\perp/D_\parallel\propto\rho^{-2}$.

We have found evidence for subdiffusive regimes $\langle \Delta
x^2\rangle \propto \Delta t^{m}$, with $m < 1$, at low enough
rigidities $\rho\lesssim10^{-2}$ and for $\eta<1$.  On analytical
grounds one expects $m=1/2$, corresponding to the so-called process of
compound diffusion~\cite{Getm63,Kirk96}, and we have found values of
$m$ close to this value indeed. However we have not been able to
investigate in detail this issue, as it is very consuming in terms of
computer time. In effect, this can be studied only using the GJ
algorithm, since it takes place at low rigidities outside the
dynamic range of the FFT algorithm. We have thus decided to postpone
the study of these anomalous regimes to a subsequent publication.

\subsection{Caracterisation of  magnetic chaos}\label{subsec.MGK}

\begin{figure}[t]
      \centering \includegraphics[width=10cm]{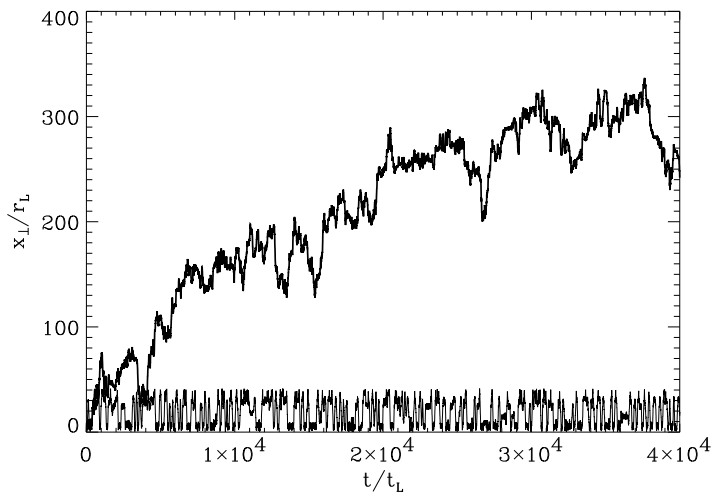}
      \caption{Transverse displacement of particles in
      three-dimensional chaotic turbulence (thick line) and in
      one-dimensional non-chaotic turbulence (thin line). In both
      cases, $\eta=0.5$, and for one-dimensional turbulence, the
      inhomogeneous component is taken to depend on the coordinate $z$,
      with the homogeneous magnetic field component lying along the $z$
      axis. Note the difference in behavior: in one-dimensional
      turbulence, the particle is confined to a flux tube and does not
      diffuse.}  \label{fig:nochaos}
\end{figure}

    When the magnetic field is a superposition of a mean field and an
irregular component depending on all three spatial coordinates, the
field line system generically exhibits chaotic solutions. For instance
it is sufficient to use a distribution of Fourier modes following a
power law in wavenumber to obtain a chaotic system. However a
two-dimensional field cannot have chaotic field lines, and a
one-dimensional system cannot produce transverse diffusion, as the
particles are confined in a flux-tube by conservation of the adiabatic
invariant~\cite{jon}. An example of this phenomenon is shown in
Fig.~\ref{fig:nochaos} in which we show the transverse wandering of a
particle in three-dimensional and one-dimensional turbulence.

    In a three-dimensional chaotic system the separation between two
initially adjacent field lines first increases exponentially $\propto
\exp(s/l_{\rm K})$ as a function of the abscissa $s$ along the field
line, with characteristic Lyapunov exponent $l_{\rm K}$, also called
the Kolmogorov length. When the separation has become larger than the
coherence length of the magnetic field, it behaves diffusively with
magnetic diffusion coefficient $D_{\rm m}\equiv \frac{\langle \Delta
r^2\rangle }{2\Delta s}$, where $\Delta r$ denotes the separation
between the two field lines.

\begin{figure}[t]
      \centering \includegraphics[width=10cm]{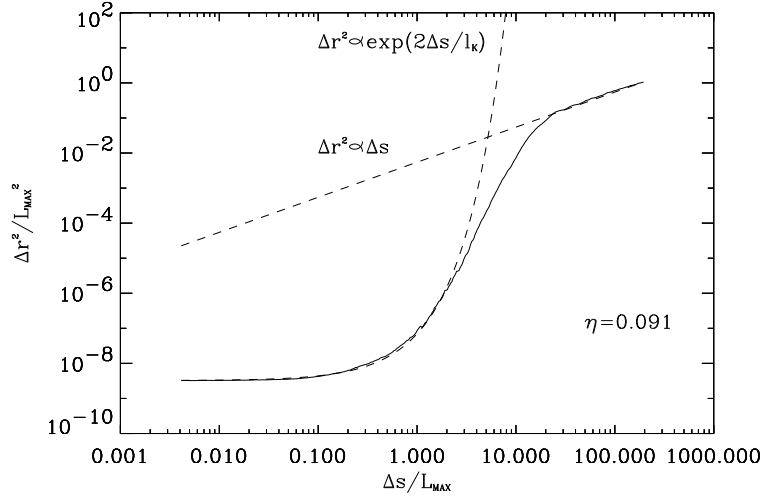}
      \caption{The square of the separation distance between two field
      lines as a function of the curvilinear abscissa along the field
      line. The exponential divergence followed by the diffusion regime
      is clearly identified. The transition between these two regimes
      occurs at $s\sim L_{\rm max}$.} \label{fig:div}
\end{figure}

    Our numerical computation of the field lines clearly displays this
two- step behaviour. In Fig.~\ref{fig:div}, we plotted the separation
squared between two field lines as a function of the curvilinear
abscissa for $\eta=0.08$. These calculations have been obtained by
integrating the equations defining the field lines, namely ${\rm
d}x/{\bar B}_x={\rm d}y/{\bar B}_y={\rm d}z/{\bar B}_z$, instead of
integrating the particle equation of motion. Figure~\ref{fig:div}
clearly shows this two-step behavior and confirms that the transition
from one regime to the other occurs when $s\sim L_{\rm max}$.

\begin{figure}[t]
      \centering \includegraphics[width=10cm,clip=true]{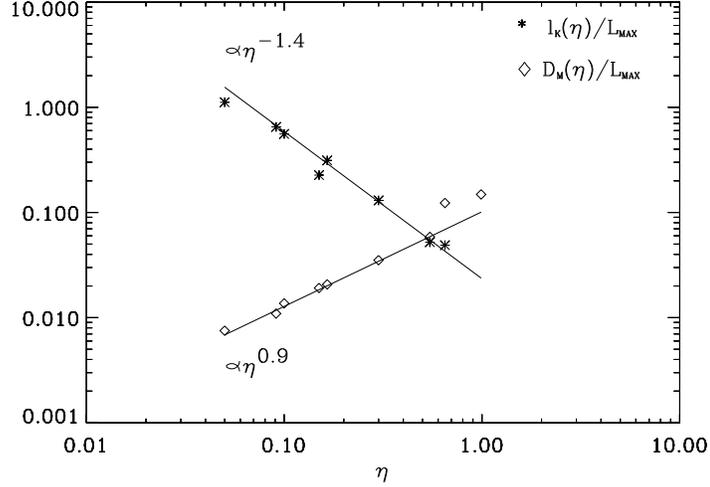}
      \caption{Kolmogorov length and magnetic diffusion coefficient as
      functions of $\eta$. The two lengths are normalized to the
      largest scale of turbulence $L_{\rm max}$.}  \label{fig:kol}
\end{figure}

This calculation allows us to measure the two lengths $l_{\rm K}$ and
$D_{\rm m}$ with a relatively good accuracy.  The results are reported as
function of $\eta$ in Fig.~(\ref{fig:kol}).

    The effective transverse diffusion of particles in a chaotic
magnetic field has been derived by Rechester \&
Rosenbluth~\cite{rero}. Here we extend their argument by assuming that
the primary transverse diffusion is anomalous (sub- or
super-diffusive).  The problem can be stated as follows.  After $n$
scattering times, parallel diffusion leads to a diffusion in
curvilinear abscissa $\langle \Delta s_{n}^2\rangle = 2D_{\parallel}
\tau_{\rm s} n$, whereas the transverse primary variation causes
transverse displacement such that $\langle \Delta x_{\perp}^2\rangle
\sim r_{\rm L}^2 n^{\alpha}$, with $\alpha = 1$ for normal diffusion, and
$\alpha <1$ for subdiffusion.  Because of field line exponential
divergence, until the separation is of order the correlation length,
say after $n_c$ scatterings, the transverse displacement is amplified
exponentially by a factor $e^{2s_{n}/l_{\rm K}}$ with
$s_{n}=\sqrt{2D_{\parallel}\tau_{\rm s}n}$. After $n_{c}$ scatterings
($n_{c} \gg 1$), an effective transverse diffusion coefficient can
then be estimated as:
\begin{equation}
D_{\perp} = \frac{\langle \Delta x_{\perp}^2\rangle }{2 \Delta
s}\left .\frac{\Delta s} {\Delta t}\right\vert_{n_{c}} \label{eq:DPKO'}
\end{equation}
Because the particles almost follow the field lines, the first factor
can be approximated by the magnetic diffusion coefficient $D_{\rm m}$,
and one gets:
\begin{equation} D_{\perp}=D_{\rm m}\frac{v}{\sqrt{3n_{c}/2}} \ .
\label{eq:DPKO"}
\end{equation}
The number $n_{c}$ is obtained by equating the separation distance and
the transverse correlation length of the field lines $l_{\perp}$ (in
our case $l_\perp\simeq L_{\rm max}$):
\begin{equation} r_{L}n_{c}^{\alpha/2}
\exp\left(\frac{\sqrt{2D_{\parallel}\tau_{\rm s}n_{c}}}
{l_{\rm K}}\right) = l_{\perp}
\label{eq:EQUAT} \end{equation}
which leads to
\begin{equation} \sqrt{n_{c}} \simeq
\frac{3}{2}\frac{l_{\rm K}}{\bar l} \log
\left[\frac{l_{\perp}}{r_{L}}\left(\frac{\bar l}{l_{\rm
K}}\right)^{\alpha}\right] , \label{eq:NC}
\end{equation}
where $\bar l \equiv v\tau_{\rm s}$ (the scattering length).  The main
result is that magnetic chaos amplifies the transverse diffusion in
such a way that it becomes a sizable fraction of the parallel
diffusion:
\begin{equation} D_{\perp} = \frac{2D_{\rm m}} {l_{\rm K}\log
\left[\frac{l_{\perp}}{r_{L}}\left(\frac{\bar l}{l_{\rm
K}}\right)^{\alpha}\right]} D_{\parallel} \label{eq:DPKO}
\end{equation}
As can be seen, the primary subdiffusion does not refrain the
effective diffusion due to chaos. When $\alpha=1$ (non-anomalous
primary diffusion), the logarithmic factor reads $\simeq
\log\left(l_\perp/g l_{\rm K}\right)$, with $g$ the scattering
function as before.

    Finally note that the above regime of diffusion applies at late
times after $n_c$ scattering times. The intermediate regime, for $n$
scattering times, with $n<n_c$, leads to sub-diffusive motion
(compound diffusion), with $\Delta x_\perp^2 \propto \Delta t^{1/2}$,
see for instance Ref.~\cite{Kirk96}. We have found evidence for such a
regime, but a detailed study of its behavior lies beyond the present
work and is defered to a later study.

  No theory gives the ratio $D_{\rm m}/l_{\rm K}$, except for some toy
models such as the Chirikov-Taylor mapping~\cite{chiri}.  However our
numerical experiment can provide a fairly accurate estimate of this
ratio. In particular we find that the Kolmogorov length $l_{\rm K}
\propto L_{\rm max}\eta^{-0.9\pm 0.1}$ and that the magnetic diffusion
coefficient $D_{\rm m} \propto L_{\rm max} \eta^{1.4\pm 0.1}$ as long
as $\eta \leq 0.5$ [see Fig.(\ref{fig:kol})]. Beyond this limit, our
calculations of $D_M$ and $l_K$ do not provide accurate estimates of
these lengths, especially for the Kolmogorov length which loses its
physical meaning when $\eta$ reaches unity.  Therefore bearing in mind
that the two diffusion coefficients become the same as $\eta
\rightarrow 1$, we conjecture that the result should be:
\begin{equation}
      D_{\perp} = \eta^{2.3\pm 0.2} D_{\parallel} \ .
      \label{eq:DPCON}
\end{equation}
\noindent when $\eta\leq 0.5$. This non perturbative result would be in
agreement with the perturbative result obtained by Chuvilgin and Ptuskin
Ref.~\cite{ptus} for small amplitude (written $A$) large scale varying
fields, the ratio between the two coefficients being proportional to $A^4$.

    Finally, it is important to note that our numerical results for
$D_\perp/D_\parallel$ shown in Fig.~\ref{fig:Dper2Dpar} have been
obtained independently of the above magnetic diffusion law. We find,
in agreement with the above relation, that $D_\perp/D_\parallel$ is
independent of $\rho$ for $\rho<1$, and that
$D_\perp/D_\parallel\propto \eta^{2.3}$ provides a good fit to the
scaling observed at $\eta < 1$. However the numerical prefactor in
this relation is rather of order $\simeq 0.2$ for $\eta <1$, whereas
it should be $\simeq1$ if the extrapolation could be taken up to
$\eta=1$. Nevertheless, the above provides solid evidence in favor of
a dominant contribution of magnetic diffusion to the process of
transverse diffusion.

\section{Some applications}

In this section, we offer revised estimates of the maximal energy that
can be attained by Fermi acceleration mechanisms by comparing the
acceleration time and the time of escape of cosmic rays outside of the
accelerating region using the results obtained in the previous
section. We first consider the case of Galactic supernovae remnants
(SNR) and so-called superbubbles, and then turn to the case of jets in
extragalactic sources.

\subsection{Supernovae remnants and super-bubbles}\label{subsec.SNRSB}

The lagging questions of the production of cosmic rays in supernovae
remnants has been recently reviewed in Ref.~\cite{kirkSN}. One of the
major problems in accounting for the observational data is that the
maximum energy achieved by the Fermi process in the SNR shock is well
below the so-called "knee"-range $10^{14}-10^{17}\,$eV. If one uses
the Bohm approximation to the diffusion coefficient, there is hope to
reach the knee energy with sufficient efficiency to expect significant
$\gamma$-ray emission resulting from $\pi^0$ decay generated by
pp-collisions. The lack of detection of these gamma-rays~\cite{aha} or
their marginal detection at best, ruined these optimistic
assumptions. The Fermi acceleration at a shock of velocity $u_{s}$ is
characterized by an acceleration timescale $t_{F1} \simeq
2D/u_{s}^2$. In most of the shock region, $D \simeq D_{\parallel}$
hence $t_{F1} \simeq \tau_{\rm s}/\beta_{s}^2 \simeq t_{\rm L}/(g
\beta_{s}^2)$, where $\beta_{s} \equiv u_{s}/c$. The maximum energy is
limited by the age of the supernovae remnants, and one thus obtains
\begin{equation}
      \epsilon_{\rm SNR} \sim 1.8 \times 10^{14} Zg
      \left(\frac{\beta_{s}}{10^{-2}}\right)^2 \left(\frac{t}{300{\rm
      yr}}\right)^2\left(\frac{B}{1\mu {\rm G}}\right) {\rm eV} \ .
      \label{eq:SNR}
\end{equation}
This result differs from \cite{kirkSN} only by the factor $3g$. This
factor is close to unity when the Bohm scaling applies; but, as we
have found in Section~\ref{subsec.NURE}, in fact
$g\propto\rho^{2/3}$. Strictly speaking this scaling is valid for
Kolmogorov turbulence, and one expects the turbulent magnetic field
downstream to differ from isotropic three-dimensional Kolmogorov, but
the above scaling serves well for order of magnitude
estimates. Moreover $\rho \ll1$ for Larmor radii smaller than
turbulence correlation length which could be the case even for the
most energetic particles. The Bohm approximation thus appears very
optimistic.

    Super-bubbles correspond to huge cavities created by $\sim100$ SNR
shock waves built around massive stars associations. The size of these
regions can be a sizable fraction of the galaxy disk thickness $h
\simeq 120\,$pc). In effect, a typical super-bubble radius can be
estimated as~\cite{pariz}

\begin{equation}
      R_{\rm SB}(t) \simeq 66\,{\rm pc}\, \left(\frac{L}{10^{38}{\rm
      erg/s}}\right)^{1/5} \left(\frac{n_{0}}{1{\rm
      cm}^{-3}}\right)^{-1/8} t_{{\rm Myr}}^{3/5}, \label{eq:RSB}
\end{equation}
where $L$ measures the mechanical luminosity of the OB-stars
association, $n_{0}$ the particle density of the surrounding
interstellar medium $\simeq 1\,$cm$^{-3}$, and $t_{\rm Myr}\sim 30$ is
the super-bubble lifetime in units of Myr.  The bubble plasma is more
dilute than the interstellar medium by at least two orders of
magnitude and thus the Alfv\'en velocity is much greater. This density
reads ~\cite{pariz}
\begin{equation}
      n_{\rm SB} \simeq 1.6 \times 10^{-2} {\rm cm}^{-3} L_{38}^{6/35}
      n_{0}^{19/35} t_{{\rm Myr}}^{-22/35} \kappa_{0}^{2/7},
      \label{eq:NSB}
\end{equation}
where $\kappa_{0}$ is a number of order unity~\cite{pariz}. The bubble
is traversed by many shock fronts propagating with velocity of order
or greater than the Alfv\'en velocity; a second order type of Fermi
acceleration is thus at work.  Its acceleration time scale is given by
$t_{F2} \sim (c/V_{A})^2 \tau_{\rm s} = (c/V_{A})^2 t_{\rm L}/g$. The
maximal energy is limited by escape of the particles, which is
governed by the diffusion across the galaxy disk thickness for the
most energetic. A strict lower limit to the time of escape $\tau_{\rm
esc}$ can be obtained by using the parallel diffusion coefficient,
since $\tau_{\rm esc}\propto 1/D$ with $D$ the diffusion
coefficient. Transverse diffusion would improve the confinement time
and thus lead to a higher maximal energy; however one should then take
into account the fact that magnetic lines come out of the galaxy disk
and unfold in the halo.  Let us consider the lower estimate:
\begin{equation}
      \tau_{\rm esc} = \frac{h^2}{2D_{\parallel}} = \frac{3h^2}{2c^2
      t_{\rm L}}g.  \label{eq:TESC}
\end{equation}
The maximal energy for acceleration by second order Fermi process in
super-bubbles then corresponds to $\tau_{\rm esc}\simeq t_{F2}$, and
reads

\begin{equation}
      \epsilon_{\rm SB} \simeq 4 \times 10^{12} {\rm eV}\, gZ
      \left(\frac{B}{1\mu \rm G}\right)^2 t_{\rm Myr}^{32/35},
      \label{eq:ESB}
\end{equation}
\noindent
where we used $n_o=1{\rm cm}^{-3}$. Therefore the second order Fermi
process in super-bubbles might cover the knee range with slightly
optimistic assumptions, since the magnetic field intensity can easily
reach $10 \mu$G in these super-bubbles, and $t_{\rm Myr}\sim
30$. Moreover, at that maximal energy the rigidity reaches unity and
therefore $g\sim 0.5 \eta$, smaller but close to unity . At this point
it is useful to recall that the confinement limiting energy of cosmic
rays in the galaxy obtained by equating the Larmor radius with the
thickness $h$ is of order $Z \times 10^{17}\,$eV.

\subsection{Extragalactic Jets and Hot Spots}\label{subsec.hotspotjet}

Extragalactic jets emanating from Active Galactic Nuclei have been
considered as possible sources of ultra high energy cosmic rays with
$E\gtrsim 10^{18}\,$eV because the confinement limit in a jet of
radius $R_{j}$ and bulk Lorentz factor $\Gamma$ is
\begin{equation}
      \epsilon_{\rm cl} = 10^{21} {\rm eV}\,
      Z\Gamma\left(\frac{B}{1{\rm G}}\right)\left(\frac{R_{j}}{1\rm pc}\right)
      \label{eq:ECL}
\end{equation}
and for the powerful, strongly collimated and with terminal hot spots
FR2 jets, the product $BR_{j}$ is roughly uniform in the jet  and is estimated
as
\begin{equation}
      BR_{j} \sim 0.1\,{\rm G.pc}\,\left(\frac{M_{*}}{10^8
      M_{\odot}}\right)^{1/2}, \label{eq:BRJ}
\end{equation}
where $M_{*}$ is the mass of the central black hole. Indeed,
asymptotically, the magnetic field at the edge of the jet is dominated
by its toroidal component; therefore the product $BR_{j}$ is governed
by the current generated by the central engine along the axis. It
slightly decreases along the jet, because the return current
progressively establishes through wrapped lines off axis like
butterfly wings~\cite{casfer,fer}. These jets are launched if the
magnetic field intensity is close to equipartition with the radiation
pressure in the central region, {\it i.e.}, within $10$ gravitational
radii. This corresponds to $B \sim 1{\rm kG}\, (M_{*}/10^8
M_{\odot})^{-1/2}$ within $10 {\rm AU}\,(M_{*}/10^8 M_{\odot})$. Thus
the performance of the jets as UHE cosmic rays accelerators tightly
depends on the nature of the central engine.  At the base of the jet,
$B\simeq100\,$mG for $R_{j} = 1\,$pc is a reasonable number. In the
hot spots of the FR2 jets such as those of Cygnus A, $B \simeq
10^{-4}\,$G for a region of size $\sim1\,$kpc. Thus with $\Gamma =
10$, the confinement condition in jets rules out the possibility of
generating ultra-high energy cosmic rays of energies larger than
$\sim10^{20}\,$eV. In the case of FR1 jets, which are less powerful,
less collimated, and without hot spots, the limiting energy is even
smaller since the product $BR_{j}$, although not well known, is very
likely lower.

As usual, the escape of the highest energy cosmic rays is governed by
diffusion across the jet and
\begin{equation}
      \tau_{\rm esc} = \frac{R_{j}^2}{2D_{\perp}} \simeq
      \frac{3}{2}\frac{R_{j}^2}{\eta^{2.3}c^2}\frac{g}{t_{\rm L}},
      \label{eq:TESCJ}
\end{equation}
where we have use our previous result that perpendicular diffusion is
governed by magnetic diffusion of the field line, and $D_\perp\sim
0.2\eta^{2.3}D_\parallel$. One thus finds that indeed most high energy
cosmic rays escape before reaching the end of the jet, since
\begin{equation}
      \tau_{\rm esc} \simeq 2\times 10^{4}{\rm
      yr}\,\frac{gZ}{\eta^{2.3}} \left(\frac{R_{j}}{1\rm pc}\right)^2
      \left(\frac{B}{1{\rm G}}\right)\left(\frac{\epsilon}{10^{18}{\rm
      eV}}\right)^{-1}, \label{eq:TESCN}
\end{equation}
to be compared with a travel time of $\simeq 1\,$Myr to travel
$300\,$kpc, the typical length of extragalactic jets. Here as well the
maximal energy for Fermi acceleration is obtained by equating
$\tau_{\rm esc}$ with the acceleration timescale for acceleration in
shocks moving at speed $\beta_s c$. This gives

\begin{equation}
      \epsilon_{\rm max} \simeq \beta_{s}\frac{g}{\eta^{1.15}}
      \epsilon_{cl}.  \label{eq:EMJ}
\end{equation}
With the plausible assumption of Kolmogorov turbulence, $g \sim 0.5
\eta \rho^{2/3}$, 
we finally obtain the maximal energy as a fraction $\beta_{s}^3$ of
the confinement energy, the estimate being weakly sensitive to the
turbulence level:
\begin{equation}
      \epsilon_{\rm max} \simeq 10^{21}{\rm eV}\,\beta_{s}^3
      Z\Gamma\left(\frac{B}{1{\rm G}}\right)\left(\frac{R_{j}}{1\rm
      pc}\right).  \label{eq:EMJ2}
\end{equation}

Centaurus A is a well-known example of active galactic nucleus,
actually the closest to us (distance 3.4~Mpc), which displays FR1 non
relativistic jets moving at speeds $\sim 5 \times 10^3\,$km/s. The
jets have both radio and X-ray band synchrotron emission with
luminosity $L_{X} \simeq 10^{39}\,$erg/s extending over several
kpc. They have been studied in detail with high resolution
interferometry~\cite{burns} and recently with the Chandra X-ray
satellite~\cite{kraft}. The radio knots and the X-ray knots are
identical in the inner jet. The minimum pressure magnetic field is
$B_{\rm eq} \simeq 60 \mu$G and the maximum Lorentz factor of the
electrons $\gamma_{\rm max} \simeq 8 \times 10^7$.  The inner jet has
radius $R_{j} \simeq 30\,$pc and a constant opening angle of
$6^\circ$. The product $BR_{j} \sim 1.8 \times 10^{-3}\,$G.pc is
clearly too low to produce ultra-high energy cosmic rays. In any case
it has been shown that even if Centaurus A could accelerate cosmic
rays to the highest energies observed $\sim 10^{20}\,$eV, their
transport to Earth, affected by diffusion according to the rules
derived in this paper would lead to strong energy losses by increased
travel distance and anisotropy incompatible with present
observations~\cite{isola}.

    In the hot spots, the escape is again dominated by diffusion at high
energies, but parallel diffusion is more likely unless there is no
ordered field. For a hot spot of size $R_{\rm hs}\sim {\rm
a\,few}\,$kpc and magnetic field intensity $B\sim10^{-4}\,$G as in
Cygnus A, the confinement limit is
\begin{equation}
      \epsilon_{\rm cl} \simeq 10^{20}{\rm eV}\,
      Z\left(\frac{B}{10^{-4}{\rm G}}\right)\left(\frac{R_{hs}}{1\rm
      kpc}\right) \label{eq:ECLHS}
\end{equation}
and the maximum energy achievable with a non-relativistic shock is
\begin{equation}
      \epsilon_{\rm max} \simeq \beta_{s}g \epsilon_{\rm cl}
      \label{eq:EMHS}
\end{equation}
The most extreme energy that can be obtained is when the turbulence is
high enough that no organised field is set up in the hot spot, and the
shock is midly relativistic $\beta_{s} \simeq 1$. But synchrotron
emission of hot spots, like those of Cygnus A, does not favor this
view. Indeed the synchrotron emission by relativistic electrons cuts
off in the infrared range. Since an electron of Lorentz factor
$\gamma$ synchrotron radiates around a frequency $\nu_{\rm syn} \simeq
115 (B_{\perp}/10^{-4}{\rm G}) \gamma^2\,$Hz, an observational upper
bound on the electron Lorentz factor is
\begin{equation}
      \gamma_{\rm max}^e \simeq 10^6
      \left(\frac{B}{10^{-4}{\rm G}}\right)^{-1/2}, \label{eq:GMAXE}
\end{equation}
and the corresponding rigidity is
\begin{equation}
      \rho_{e} \equiv \frac{2\pi r_{\rm L}}{R_{\rm hs}} \simeq 0.3 \times
      10^{-7}\left(\frac{B}{10^{-4}{\rm G}}\right)^{-3/2}
      \left(\frac{R_{\rm hs}}{1{\rm kpc}}\right)^{-1}.
\label{eq:ROE}
\end{equation}
Now this maximum electron energy is obtained by the same Fermi
acceleration process, limited by synchrotron losses. We remind that
the characteristic time for a synchrotron radiative process of an
electron of energy $\epsilon=\gamma m_ec^2$ is
\begin{equation}
t_{\rm syn} = \frac{6\mu_o\epsilon}{4\sigma_TcB^2\gamma^2}
\label{eq:Synch}
\end{equation}
\noindent where $\sigma_T$ is the Thomson cross-section and $\mu_o$
the magnetic permitivitty of vacuum. By equating this loss time with
the first-order Fermi acceleration timescale, we obtain the maximum
Lorentz factor that can be achieved
\begin{equation}
      \gamma_{\rm max}^e \simeq 10^{10} \beta_{s} g^{1/2}
      \left(\frac{B}{10^{-4}{\rm G}}\right)^{-1/2}, \label{eq:GMAXE2}
\end{equation}
and the scattering function is calculated for the maximum rigidity
$\rho_e$.
Therefore, assuming again Kolmogorov turbulence with $g(\rho_{e})
\simeq 0.5 \eta \rho_{e}^{2/3}$, we obtain the approximate value of
the turbulence level by equating the two expressions for $\gamma_{\rm
max}^e$:

\begin{equation}
      \eta \simeq 0.2
      \left(\frac{\beta_{s}}{0.1}\right)^{-2}\left(\frac{B}{10^{-4}\rm
      G}\right) \left(\frac{R_{\rm hs}}{1\rm kpc}\right)^{2/3}.
      \label{eq:TURL}
\end{equation}
Since $\beta_{s}>0.1$ is very likely, the required turbulence level is
rather low. This, in turn, reduces drastically the maximum energy of
cosmic ray acceleration, using Eq.~(\ref{eq:EMHS}), Eq.~(\ref{eq:ECLHS})
and knowing that $g\sim\eta\rho^{\beta-1}$:
\begin{equation}
      \epsilon_{\rm max} \simeq (\eta \beta_{s})^{1/(2-\beta)}
      Z\left(\frac{B}{10^{-4}\rm G}\right) \left(\frac{R_{\rm hs}}{1\rm
      kpc}\right) 10^{20}{\rm eV}\ , \label{eq:EMHS2}
\end{equation}
where we explicited the scaling with $\beta$ the exponent of the power
spectrum of magnetic fluctuations; however note that the estimate of
$\eta$ must be changed with $\beta$. For Kolmogorov turbulence,
$\beta=5/3$ and using the upper bound on $\eta$ Eq.~(\ref{eq:TURL})
above, the prefactor is of order $10^{-6}(\beta_s/0.1)^{-3}$, and
acceleration is not sufficient to account for the highest energy
cosmic rays by several orders of magnitude.

      This limit cannot be circumvented easily, since it is severely
constrained by the cut-off frequency of the electrons synchrotron
emission. The only parameter that could be modified without affecting
this cut-off frequency is the turbulence index $\beta$.  If one
considers Kraichnan turbulence $\beta = 3/2$ instead of Kolmogorov
turbulence, the prefactor $(\beta_{s} \eta)^3$ is changed into
$(\beta_{s} \eta)^2$, but $\eta$ itself is lowered by a factor $10$
due to the modified dependence of $g$ on $\rho_e$. Furthermore if hot
spots were to radiate synchrotron emission in X-rays, this would
increase $\gamma_{max}^e$ by a factor $10$ only, and would not affect
drastically our conclusions. Finally the numbers considered above are
consistent with recent observations of the Cygnus A hot spots by
Chandra~\cite{wilson}, which give an accurate measurement of the
magnetic field intensity $\sim 1.5 \times 10^{-4}\,$G to within a few
tens of percents, as obtained by the ratio of the synchrotron-self
Compton luminosity over the synchrotron luminosity, a value which is
furthermore close to the equipartition value if there are no protons!
These measurements also confirm model dependent estimates proposed in
1986~\cite{pelro}. Finally, as an aside, the same reasoning allows to
estimate the level of turbulence required to get the X-ray emission in
Centaurus A: with $\gamma_{max} \simeq 8 \times 10^7$, $\rho_{e}
\simeq 1.2 \times 10^{-4}$, and $\eta \simeq 10^{-2}/\beta_{s}^2$.

\section{Conclusion}\label{sec.DISCU}

Let us first summarize the results we have obtained. The scattering
function $g$ has been found to follow the scaling predicted by
quasi-linear theory in the inertial range $\rho_{\rm min}<\rho < 1$
for weak to strong turbulence.  However we found that scattering still
operates for $\rho < \rho_{\rm min}$, contrary to the predicted sudden
drop of the scattering function; this facilitates the injection of
particles in Fermi processes.  For Larmor radii larger than the
correlation length $\rho>1$, scattering decreases as a power-law in
rigidity unlike the predicted sudden drop of $g$. Therefore high
rigidity particles still diffuse. One should also mention that the
lack of scattering encountered in weak turbulence theory for particles
having pitch angle close to $90^0$ is cured in strong enough
turbulence.
 
    The perpendicular diffusion turns out to be very different from
the prediction of the quasi-linear theory. Our investigation of the
chaos of magnetic field lines characterised by a Kolmogorov length and
a diffusion coefficient with space increment indicates that this
process of magnetic diffusion governs the transverse diffusion of
particles.

    Our numerical experiment shows that the phenomenological Bohm
approximation, characterized here by $g \sim 0.5$ and $D = \alpha_{B}
r_{L}v$ with $\alpha_{\rm B}\sim 0.7$, only applies in a limited range
of rigidities $0.1\lesssim\rho\lesssim1.$, and only in the case of
pure turbulence $\eta=1$. Many estimates in astroparticle physics,
that rely on the Bohm conjecture, must be reconsidered.

    The slow decrease $g \propto \rho^{-4/3}$ of scattering for cosmic
rays with Larmor radius larger than the correlation length of the
magnetic field, which implies $D \propto \rho^{7/3}$, is of potential
importance to the transport of high energy cosmic rays in our Galaxy
as well as ultra-high energy cosmic rays in the intergalactic medium.

    The accurate knowledge of the transport coefficients allows to be
more conclusive than before on the performances of Fermi acceleration
in some astronomical sources of high energy cosmic rays such as
supernovae remnants, super-bubbles and extragalactic jets.  Using new
Chandra data, the turbulence level and the maximum energy for
electrons and for cosmic rays can be determined. We confirm the
difficulty to obtain energies larger than $10^{13}\,$eV in supernovae
remnants and shows that the "knee" range of the cosmic ray spectrum
could be accounted for by second order Fermi acceleration in super
bubbles. We also confirm that FR1 jets, such as Centaurus A, although
radiating synchrotron in X-rays, cannot produce UHE Cosmic rays. On
the contrary, FR2 jets can produce cosmic rays up to $10^{20}\,$eV,
but presumably not more, owing to a fairly good confinement; however
most high energy cosmic rays escape before reaching the end of the
jet. Hot spots of powerful radio-galaxies have always been considered
as a promising source, but we have found that, because of the low
turbulence level implied by the synchrotron cut off frequency, cosmic
rays escape rapidly along the mean field lines by fast parallel
diffusion and acceleration is not effective above $\epsilon_{\rm
max}\sim 10^{14}\,$eV for a shock velocity $\beta_{s} \simeq 0.1$.

Our paper left opened several important issues that we are currently
investigating. In particular it seems crucial to investigate in more
detail the existence of subdiffusive regimes at low ridigities for
which we have found evidence. These regimes play a crucial role in the
acceleration processes at perpendicular
shocks~\cite{Kirk96}. Furthermore, we have described magnetic
turbulence as an ensemble of magneto-static modes distributed
according to a power law spectrum. This approximation is justified by
the small Alfv\'en velocity when compared to the velocity of the
particles.  However it would be interesting to investigate the effect
of temporal and spacial intermittency on the transport
properties. Finally we are currently investigating the transport
properties of particles in non-isotropic turbulence as may be
encountered in the vicinity of a shock wave, in particular in the
downstream medium. The consequences on Fermi acceleration will be
presented in a forthcoming paper.

\appendix

\section{Theoretical approach}\label{sec.appen}

The diffusion resulting from the random variations of the momentum due
to the irregular magnetic field can be formalized as follows.  Energy
conservation, and thus $p$ conservation, allows to treat the problem
as random rotations of the unit vector ${\bf u}$ such that ${\bf p} =
p {\bf u}$: ${\bf u}(t) = R(t,t_{0}) {\bf u}(t_{0})$.  Assuming that
the correlation functions of the components of ${\bf u}$ are
integrable over a caracteristic time $\tau_{\rm s}$, the diffusion
coefficients are given by
\begin{equation}
      D_{ij}=v^2 \int_{0}^{\infty} \langle 
u_{i}(t)u_{j}(t+\tau)\rangle  {\rm d}\tau \ .
      \label{eq:A1}
\end{equation}
The correlation matrix is derived by making the appropriate average,
after solving the stochastic equation:
\begin{equation}
      \dot {\bf u} = \Omega(t) \bf u
      \label{eq:A2}
\end{equation}
where the gyro-matrix $\Omega(t) = {\rm sign}(q)
\sum_{\alpha}b_{\alpha}(t)J_{\alpha}$, $b_{\alpha}(t)$ being the
reduced components of the magnetic field experienced by the wandering
particle and $J_\alpha$ is a $3\times 3$ matrix, with components
$(J_i)_{jk}= \epsilon_{ijk}$, where $\epsilon_{ijk}$ is the fully
anti-symmetric Levi-Civita tensor, and $\epsilon_{123}\equiv 1$.

Note that the $J_{\alpha}$ are generators of a Lie algebra such that
\begin{equation}
      J_{\alpha}J_{\beta} = {\bf e}_{\beta} \otimes {\bf e}_{\alpha} -
      \delta_{\alpha \beta}I_{d}
      \label{eq:A3}
\end{equation}
where $I_{d}$ represents the identity matrix, $J_{\alpha}^2 =
-\Pi_{\alpha}^{\perp}$ and $J_{\alpha}J_{\beta}-J_{\beta}J_{\alpha} =
\varepsilon_{\alpha \beta \gamma}J_{\gamma}$, where $\{ {\bf
e}_{\alpha} \}$ is the orthonormal basis, $\Pi_{\alpha}^{\perp}$ the
orthogonal projector over the plane transverse to the direction
$\alpha$.  We have $b_{1}=\tilde b_{1}$, $b_{2}=\tilde b_{2}$ and
$b_{3}=b_{0}+\tilde b_{3}$, with $\langle \tilde {\bf b}^2\rangle
=\eta$ and $b_{0}^2 =1-\eta$.  Moreover $\tilde {\bf b}(t) = \tilde
{\bf b}[{\bf x}_{0}+\rho {\bf \xi}(t)]$, with $\dot {\bf \xi} = {\bf
u}$. The time variable is measured in Larmor time units, the space
variables are reduced to $L_{\rm max}$ and wave numbers are
accordingly dimensionless and varies from $1$ to $1/\rho_{m}$, where
$\rho_m\equiv k_{\rm min}/k_{\rm max}=L_{\rm min}/L_{\rm max}$.  We
make the three following assumptions:
\begin{itemize}
      \item[i)]{The random process becomes stationary beyond the
      correlation time $\tau_{\rm s}$.}

      \item[ii)]{The random process can be approximated by a Markoff
      process beyond the integration time $\tau_{\rm s}$.}

      \item[iii)]{The random process $\tilde {\bf b}(t)$ is supposed to be
      specified. For instance, it is gaussian with a known correlation
      function $\langle  \tilde {\bf b}(t).\tilde {\bf b}(t')\rangle 
\equiv \Gamma(t-t')$.}
\end{itemize}
Assumptions i) and ii) allow to calculate the correlation function
with an average matrix that describes the relaxation of the
correlations:
\begin{equation}
      \langle  u_{i}(t)u_{j}(t+\tau)\rangle  = \bar 
R_{ij}(\tau)\langle  u_{i}(t)u_{j}(t)\rangle
      \label{eq:A4}
\end{equation}
where $\bar R_{ij}(\tau) = \langle R_{ij}(t+\tau,t)\rangle $.
Assumption iii) is not exact, of course; however the numerical
experiments provide correlation functions that allow to get a good
"guess". Then the theoretical method allows to calculate the solution
through iterations, starting with a gaussian approximation, and then
estimating, if necessary, non-gaussian corrections, given a skewness
factor.

The formal solution reads:
\begin{equation}
      \bar R(t,t_{0}) = \left\langle T \exp\left[\int_{t_{0}}^t {\rm d}\tau
\Omega(\tau)\right]\right\rangle \ , \label{eq:A4a}
\end{equation}
where the symbol $T$ represents the "time ordering operator" that
organises the expansion of the exponential operator in products of non
commutative operators that are in chronological order. The result can
be factorized as the product of the unperturbed rotation in the mean
field times some relaxation operator:
\begin{equation}
      \bar R(t,t_{0}) = R_{0}(t-t_{0}).\left\langle T
      \exp\left[\int_{t_{0}}^t {\rm d}\tau \tilde
      \Omega(\tau)\right]\right\rangle \ , \label{eq:A5}
\end{equation}
with $\tilde \Omega(t) \equiv R_{0}^{-1}(t-t_{0}) \delta \Omega(t)
R_{0}(t-t_{0})$ (see \cite{kubo,frisch,pello} for technical details).

\subsection{Quasi-linear approximation $\eta \ll 1$}\label{subsec.smalleta}

For $\eta$ small and for a broad magnetic spectrum insuring a short
correlation time of the random force compared to the scattering time,
the quasi linear theory applies~\cite{jok}. This allows to make two
approximations. First, the relaxation operator can be calculated to
the lowest order, the so-called "Bouret
approximation"~\cite{frisch,pello}, which corresponds to a summation
of all the "unconnected diagrams":
\begin{equation}
      \bar R(t) = e^{(\Omega_{0}+M)t}
      \label{eq:A6}
\end{equation}
with
\begin{equation}
      M = \int_{0}^{\infty} {\rm d}\tau \sum_{\alpha}\left\langle \tilde
      b_{\alpha}(t) \tilde b_{\alpha}(t-\tau)\right\rangle
      J_{\alpha}R_{0}(\tau)J_{\alpha} \label{eq:A7}
\end{equation}
The simplest way to derive this result is to linearize the evolution
equation for $R(t,t_{0})$:
\begin{equation}
      \frac{d}{dt}\delta R = \Omega_{0} \delta R + \delta \Omega \bar
      R + \ldots \ ;
      \label{eq:DELR}
\end{equation}
which one then solves to lowest order for $\delta R$ as a function of
$\delta \Omega$ and inserts the result in the evolution equation for
$\bar R$.  Then we use the isotropy of the spectrum to obtain
\begin{equation}
      M = \frac{1}{3}\int_{0}^{\infty} {\rm d}\tau \Gamma(\tau)
      \sum_{\alpha}J_{\alpha}R_{0}(\tau)J_{\alpha} \ .
      \label{eq:A8}
\end{equation}
The sum of operators can be simplified to:
\begin{equation}
      \sum_{\alpha}J_{\alpha}R_{0}(\tau)J_{\alpha}=
      -2\cos \omega_{0} \tau \Pi_{\parallel} -(1+\cos \omega_{0} \tau)
      \Pi_{\perp} - \sin \omega_{0} \tau J_{3} \ ,
      \label{eq:AA8}
\end{equation}
where $\omega_{0}$ is the reduced Larmor pulsation in the mean field,
thus $\omega_{0} = \sqrt{1-\eta}$.

Second, the correlation function of the magnetic irregularities
experienced by the particles is calculated with unperturbed
trajectories.
\begin{equation}
      \Gamma(\tau) \simeq \Gamma_{0} (\tau) = \int \frac{{\rm d}^3k}{(2\pi)^3}
      S_{3D}({\bf k}) e^{i\rho{\bf k.\xi}_{0}(\tau)}
      \label{eq:A9}
\end{equation}
where ${\bf \xi}_{0}(\tau) = \int_{0}^{\tau} {\rm d}\tau' R_{0}(\tau').{\bf
u}(0)$, and $S_{3D}$, as the notation indicates is the
three-dimensional power spectrum in Fourier space.  These two
calculations, thanks to commutation properties, lead to a matrix of
the form
\begin{equation}
      \bar R(t) = R_{0}(t) \exp\left[-g_{\parallel}\Pi_{\parallel}t-
      g_{\perp}\Pi_{\perp}t-g_{0}J_{3}t\right]
      \label{eq:AB}
\end{equation}
The factors $g$ are small numbers of order $\eta$ that contrains the
usual resonances of the quasi linear theory in $\pi
\delta(k_{\parallel}\rho \vert \mu \vert -n \omega_{0})$. These
resonances come from the cosine and sine factors in Eq.~(\ref{eq:AA8})
and from the expansion into a Fourier sequence in $n\omega_{0}$ of the
exponential involved in $\Gamma_{0}(\tau)$, see Eq.~(\ref{eq:A9}),
which introduces Bessel functions of all orders. However, in pratice,
only the main resonances for $n=\pm 1$ are retained because the higher
resonances involve shorter and shorter wavelengths which contain less
and less energy for usual power law spectra. The contribution in
$J_{3}$ modifies the gyro-pulsation in the rotation matrix $R_{0}(t)$
and therefore is unimportant. We finally retain the following result:
\begin{equation}
      \bar R(t) = e^{-g_{\parallel}t}\Pi_{\parallel} +
      e^{-g_{\perp}t}R_{0}^{\perp}(t) \ ,
      \label{eq:AA}
\end{equation}
where $R_{0}^{\perp}(t)$ is the product of the rotation and the
transverse projector.  We thus obtain the unexpected result that the
transverse relaxation is longer than the parallel relaxation since
$g_{\parallel}=2g_{\perp}$.  Then the correlation functions are
obtained by averaging over ${\bf u}(0)$ and there comes a major
problem of quasi-linear theory because the functions $g$ are
proportional to $\eta (\rho \vert \mu \vert)^{\beta -1}$ for $\vert
\mu \vert > \mu_{m} \equiv \rho_{m}/\rho$ and vanish for $\vert \mu
\vert \leq \mu_{m}$ because of the lack of resonance. This introduce
long tail contributions to the correlation functions. This is the
symptom of the "sticky" regime for pitch angles close to $90^0$ that
tends to dominate the diffusion coefficients; which requires to take
into account mirroring effects and/or overlapping of the resonances
for $\mu>0$ close to $\mu_{m}$ and those for $\mu<0$ close to
$-\mu_{m}$, as suggested in~\cite{pel99}. This difficulty disappears
in strong turbulence and for large enough Larmor radii.

\subsection{Theoretical hints with no mean field}\label{subsec.nomfh}

The fully deductive theory of this regime is quite difficult.  However
some attempt can be proposed in the case where $g < 1$, i.e.  $t_{\rm
L} < \tau_{\rm s}$, when the correlation time [decay time of
$\Gamma(\tau)$] is shorter than the scattering time. Thus, for a time
longer than the correlation time, we can keep part of the quasi-linear
theory, namely the expression of the relaxation operator involving the
integral over $\Gamma(\tau)$. Technically, this corresponds to the
summation of the unconnected diagrams of the expansion of $\bar
R(t,t_{0})$ in Eq.~(\ref{eq:A4a}), the other diagrams ("nested" and
"crossed") being of smaller orders. Therefore the correlation
functions $C_{ij}(t)$ asymptotically decay like $e^{-g_{*}t}$ and
\begin{equation}
      g \simeq g_{*} \simeq \frac{2}{3}\int_{0}^{\infty} \Gamma(\tau)
      {\rm d}\tau \ .
      \label{eq:gnomf}
\end{equation}
Now the main difference comes from the estimation of the
correlation function of the field experienced by the particles:
\begin{equation}
      \Gamma(\tau) \simeq \int \frac{{\rm d}^3k}{(2\pi)^3}
      S_{3D}({\bf k}) \langle  e^{i\rho{\bf k.\xi}(\tau)}\rangle  \ .
      \label{eq:gamm}
\end{equation}
We propose the following heuristic estimate. Because the particles
follow the field lines when their Larmor radius is smaller than the
wavelength of the modes, we consider only the modes such that
$k\rho > 1$. The dominant contribution in the averaged exponential is
then for short time, ${\bf \xi}(\tau) \simeq \tau {\bf u}(0)$.
Because of the random distribution of ${\bf u}(0)$ over the unitary
sphere, we get
\begin{equation}
      \Gamma(\tau) \simeq \int_{k\rho>1}S(k)\frac{\sin k\rho \tau}{k\rho
      \tau}{\rm dk} \ .
      \label{eq:gammaf}
\end{equation}
A similar result is obtained with a gaussian evaluation of the
average
\begin{equation}
      \langle  e^{i\rho {\bf k.\xi}(t)}\rangle  = 
\exp\left[-\frac{1}{3}k^2\rho^2 t\int_{0}^t
      C(\tau){\rm d}\tau\right] \ .
      \label{eq:A14}
\end{equation}
Inserted into the integral over the spectrum (restricted to $k\geq
1/\rho$), it leads to:
\begin{equation}
      \Gamma(\tau) \simeq \int_{k>1/\rho} {\rm d}k S(k)
      \exp\left[-\frac{1}{3}k^2\rho^2 \tau \int_{0}^{\tau} 
C(\tau'){\rm d}\tau'\right] \ .
      \label{eq:A15}
\end{equation}
In the integral, $\int_{0}^{\tau} C(\tau'){\rm d}\tau'$ can be
approximated by $\tau$ for $\tau < \tau_{\rm s}$ ($=1/g$ in reduced units)
and by $1/g$ for $\tau > \tau_{\rm s}$. Therefore
\begin{equation}
      g \simeq \frac{2}{3}\int_{k>1/\rho} {\rm d}k
      S(k)\left[\frac{\sqrt{3\pi}}{2k\rho}
      \Phi\left(\frac{k\rho}{g}\right) +
      \frac{3g}{k^2\rho^2}e^{-\frac{k^2\rho^2}{3g^2}}\right] \ ,
      \label{eq:A16}
\end{equation}
where $\Phi (x) = \frac{2}{\sqrt{\pi}}\int_{0}^x e^{-y^2}{\rm d}y$.
When $g$ is small, because $k\rho \geq 1$, we get a simple result
close to the previous one:
\begin{equation}
      g \simeq \sqrt{\frac{\pi}{3}}\int_{k>1/\rho} {\rm d}k
      \frac{S(k)}{k\rho} \ .
      \label{eq:A17}
\end{equation}
This gaussian evaluation indicates the error made by the previous
assumption.  Thus, for small $\rho$, we obtain the extension of the
quasi-linear result, namely $g \sim \rho^{\beta -1}$, and for $\rho
>1$ to $g \sim 1/\rho$. These two approximations are in agreement with
the numerical experiments, except that the measured drop is in
$\rho^{-1.3}$ instead of $\rho^{-1}$ here. Now the range of $\rho$
values where $g$ is on the order of unity corresponds to the "Bohm
estimate", which is, in fact, the maximum value of $g$ achieved for
$\rho \sim 1$ only.  We thus propose the following final estimate for
the scattering function:
\begin{equation}
      g \simeq \frac{\pi}{3} \int_{k\rho>1}\frac{S(k)}{k\rho}{\rm d}k \ .
      \label{eq:gf}
\end{equation}
The error on the coefficient is of order ten percent.

\end{document}